\documentclass[12pt]{article}

\usepackage{graphicx}
\usepackage{dcolumn}
\usepackage{bm}
\usepackage{amsmath}
\usepackage{amssymb}
\usepackage{verbatim}
\usepackage{color}
\usepackage{esvect}

\usepackage[normalem]{ulem}

\usepackage{hyperref}
\hypersetup{
    colorlinks=true,
    linkcolor=blue,
    filecolor=magenta,      
    urlcolor=cyan,
}

\begin{document}

\title{Before and beyond the Wilson-Cowan equations}
\author{Carson C. Chow and Yahya Karimipanah\\
Laboratory of Biological Modeling, NIDDK, NIH}
\date{\today}
\maketitle

\begin{abstract}
The Wilson-Cowan equations represent a landmark in the history of computational neuroscience. Among the insights Wilson and Cowan offered for neuroscience, they crystallized an approach to modeling neural dynamics and brain function. Although their iconic equations are used in various guises today, the ideas that led to their formulation and the relationship to other approaches are not well known.  Here, we give a little context to some of the biological and theoretical concepts that lead to the Wilson-Cowan equations and discuss how to extend beyond them.

\end{abstract}

\section{Introduction}\label{sec:intro}

Wilson and Cowan published the first of two classic papers in 1972 \cite{wilson_excitatory_1972}, wherein they obtained a set of coupled mean field equations providing a coarse-grained description of the dynamics of a network of excitatory and inhibitory neuron populations. The second paper in 1973 included spatial dependence \cite{wilson_mathematical_1973}.  Their work provided a demarcation in the practice of theoretical neuroscience. Prior to Wilson and Cowan, the nascent field of computational neuroscience was still grappling with how best to model neural dynamics. Thereafter,
it began to coalesce onto a common set of neural network equations and studying their properties became the focus. 

By the mid twentieth century, it became an established idea that at least some information processing in the brain is performed via global (population) activity (as opposed to the single neuron level) \cite{hosokawa1958sholl}. A number of studies looked into the dynamics of neuronal populations in order to find a theoretical framework for studying the collective behavior of neurons \cite{beurle1956properties,griffith_field_1963,griffith_field_1965}.  Given the complex nature of the brain and our lack of access to details of the microscopic processes, building a dynamical theory naturally called for the use of statistical approaches, including coarse-graining and mean field theory.

Course-graining considers the system at a lower resolution, thereby reducing the degrees of freedom of the system. However, some information will be lost. The strategy is to ensure that the lost information is not crucial to understanding the phenomenon of interest. 
The classic example is the dynamics of gas molecules in a closed room. If we are only interested in whether the room will be comfortable, then the coarse-grained measures of temperature, pressure, and relative densities of the constituent molecules (e.g. water, nitrogen, oxygen, carbon dioxide, etc.) is sufficient. The actual microscopic dynamics of the over $10^{23}$ molecules is irrelevant. However, in systems that are highly coupled, such as the brain, it is not clear what information is important. Mean field theory is a coarse-graining approach that captures the relevant mean dynamics while discounting (hopefully irrelevant) fluctuations around the mean. Mean field theory was originally developed for magnetism \cite{le2004equilibrium,peliti2011statistical}, which involves the alignment of atomic spins under the competing effects of coupling between the spin magnetic moments of each atom and disorder quantified by the temperature. Reminiscent of neurons in the brain, the dynamics are highly complex since a perturbation of one spin may lead to unending reverberating loops where each spin influences other spins which in turn affect themselves. Mean field theory cuts through this complexity by imposing self-consistency between the role of a spin as the influencer of other spins and the one being influenced by other spins. More concretely, the mean field equations are derived by computing the mean state of one spin given fixed yet unknown values of all the coupled spins then equating the values of the other spins to that of the first spin. If the number of coupled spins is large enough and the spins are uncorrelated enough then the variance of the spin state will go to zero (by the central limit theorem) and the mean will be the only relevant quantity. The neglect of correlations and higher order statistics is the hallmark of mean field theory.

The original Wilson-Cowan equations were a coarse grained mean field system for a continuous activity variable representing the proportion of a local population of neurons that is firing or active at any moment of time. However, this interpretation is not rigidly adhered to and often the activity variable is deemed a physical quantity like the mean firing rate or an electrochemical potential of a single neuron or small group of neurons.
The Wilson-Cowan and related equations became ubiquitous because they were able to model many elements of observed neural phenomena. Here we present a brief (and grossly incomplete) review of the history of some key ideas that led to their inception. We discuss when the assumptions of the equations are valid, when they break down, and how to go beyond them to take into account of fluctuations and correlations. 

\section{What are the Wilson-Cowan equations?}\label{sec:equations}

The use of neural network equations is pervasive. They have been referred to as Wilson-Cowan equations \cite{wilson_excitatory_1972,cowan_wilsoncowan_2016,Kilpatrick2013}, Amari equations \cite{Potthast2013}, Cohen-Grossberg equations \cite{lu2003new}, rate equations \cite{Sompolinsky1988,shriki_rate_2003,vogels2005neural}, graded-response equations \cite{hopfield_neurons_1984,hopfield_computing_1986}, continuous short term memory (STM) equations \cite{grossberg1969some,cohen_absolute_1983,grossberg1988nonlinear}, and neural field equations \cite{bressloff2011spatiotemporal,coombes_waves_2005}. Here, we will adopt the term activity equations.
Most of these equations come in one of three forms.
The actual activity equations in Wilson and Cowan's 1972 paper are
\begin{align}
\tau \dot{a_i} &= -a_i + (1- ra_i)f_j\left(\sum_j w_{ij} a_j + I_i \right)
\label{eqn:WC_format}
\end{align}
where $a_i$ is neural activity of a local population indexed by $i$, $f_i$ is a nonlinear response or gain function, $w_{ij}$ is a coupling weight matrix, $I_i$ is an external input, $\tau$ is a decay time constant, and $r$ is a refractory time constant. Wilson and Cowan explicitly separated the effects of excitatory and an inhibitory neurons. Their 1973 paper introduced a continuum version:
\begin{align*}
\tau \dot{a}(x) &= -a(x) + \big(1- ra(x)\big) \, f\left( \int_\Omega w(x-y) a(y) dy + I(y) \right)
\end{align*}
where $x$ is a spatial variable on a domain $\Omega$.
The other two common forms are 
\begin{align}
    \tau \dot{s_i} &= -s_i + \sum_j W_{ij} f_j(s_j) + I_i \label{eqn:Amari_format}\\
\tau \dot{s_i} &= -s_i + (B_i - s_i)\sum_j W_{ij} f_j(s_j) + I_i\label{eqn:CG_format}
\end{align}
which Grossberg~\cite{grossberg1988nonlinear} calls the additive and shunting short term memory (STM) equations respectively.  Grossberg and colleagues use (\ref{eqn:Amari_format}) and (\ref{eqn:CG_format}) with excitatory and inhibitory effects separated.
Amari and subsequent neural field papers use the continuum version of (\ref{eqn:Amari_format}), often with excitation and inhibition lumped together so neurons no longer obey Dale's law but simplifies the analysis.

The term Wilson-Cowan equations often refer to either (\ref{eqn:WC_format}) with $r=0$ or (\ref{eqn:Amari_format}). For constant external input, these two equations are actually identical \cite{grossberg1988nonlinear} as can be seen by setting
$s_i = I_i+\sum_j w_{ij} a_j$, and taking the derivative: 
\begin{align}
\dot{s_i} &= \sum_j w_{ij}\dot{a_j}=\sum_j w_{ij}\left[-a_i + f_j\left(\sum_j w_{ij} a_j + I_i \right)\right]\\
&= - s_i  + \sum_{j} w_{ij} f_j\left(s_j \right) +I_i
\end{align}
From this perspective, if $a$ is neural activity then $s$ is best viewed as a synaptic input or drive \cite{pinto_quantitative_1996}.
We will use the term Wilson-Cowan (WC) equations to exclusively refer to (\ref{eqn:WC_format}). For (\ref{eqn:Amari_format}) we will use the term graded-response model. For a quasi-stationary state where $s_i \approx \sum_{j} w_{ij} f_j\left(s_j \right) +I_i$, $a_i$ can be interpreted as a neuronal mean firing rate 
\begin{align}
a_i = f_i(s_i)
\label{eqn:gain}
\end{align}
and $f_i(s)$ is a single-neuron gain function (FI curve) specifying rate as a function of input. Equation (\ref{eqn:WC_format}) would then be considered a rate-based model equivalent to the graded-response model  (\ref{eqn:Amari_format}).

\section{Before the Wilson-Cowan equations}\label{sec:BWC}

There are three essential ingredients that activity equations include: 1) a continuous time dependent activity variable, 2) a linear weighted sum over these variables in the input, and 3) a nonlinear gain or activation function linking input to output. Underlying these ingredients are coarse-graining and mean field theory. Here we explore some of the history behind these ideas. A list of milestones is in the box.

It was well known throughout the last century that neurons fired action potentials if given sufficient stimulus. Thus it was reasonable to model them as binary state machines that activate when the inputs exceed a threshold.
In 1943, McCulloch and Pitts \cite{mcculloch1943logical} showed that a network of such binary threshold neurons with both excitatory and inhibitory connections is capable of producing any logical operation and thus performing universal computations. This launched the study of neural networks in computer science, artificial intelligence, and neuroscience.

Shortly thereafter a dissenting opinion emerged that sprouted a branch of research that would eventually lead to the Wilson-Cowan equations and dynamical systems modeling of the brain. Shimbel and Rapoport \cite{shimbel_statistical_1948} argued that the McCulloch and Pitts ``neural net" approach fell short of building plausible theories for neural systems. They argued that countless numbers of equivalent networks could produce the same desired output and that neural nets were not robust to the failure of a few neurons. Additionally, it was unlikely that genes predetermined the details of the microscopic structure of biological systems but rather imposed statistical traces on the macroscopic properties. They proposed that the goal should be to seek the statistical properties that govern the evolution and function of a biological system, rather than proposing specific network connectivity to perform a particular task. 

They derived a dynamic equation governing the firing probability of neurons located in a given region of a brain network in terms of a recursive map. Like McCulloch and Pitts they assumed binary neurons and supposed that a particular synapse receives input from $n$ axon terminals, which they called ``bulbs" partitioned into $p$ ``bulb groups". The probability that a neuron will fire depends on an integer number of bulbs exceeding a threshold $h$. Neuron firing dynamics are decomposed into a set of probabilities governing the connectivity and synaptic properties.
The fraction of neurons at location $Q$ firing at time $t=1$ was presumed to be
\begin{equation}
f_{1}(Q)=\sum_{n, p, h} P(Q; n,p,h) P_{n p h}[I_{0}(Q) ]
I_{0}(Q)=\int_{\Omega} O(Q, X) f_{0}(X) d X
\end{equation}
where $f_0(X)$ is the firing probability of neurons at $X$ on a domain $\Omega$ at time $t=0$, $O(Q, X)$ is the probability that a bulb group at $Q$ originates from $X$, $P(Q; n,p,h)$ is the probability that a neuron at $Q$ has a firing threshold $h$ and synapses of type $(n,p)$, and  
$ P_{n p h}[I_{0}(Q)]$ denotes the conditional probability that an active neuron at $Q$ is of $(n,p,h)$ type. A crucial idea for these equations is that the probability of firing at a given time only depends on the probability of firing in the previous time and not on any higher order statistic. This is an implicit mean field theory assumption. Although $P_{n p h}[I_{0}(Q)]$ are derivable in principle, it would be a daunting task because the probability of activation depended on the microscopic details of synaptic configurations.

Beurle \cite{beurle1956properties} introduced coarse-graining, which resolved the synaptic structure problem.  He was inspired by experimental findings on the distribution of dendritic fibers as a function of distance from the cell body \cite{sholl1955organization,uttley1955probability}.  Anatomical studies had started to focus on connectivity patterns of neuronal aggregates (as opposed to the connections of individual neurons), providing probability distributions instead of specific circuits \cite{feldman_large-scale_1975}.  Berule's coarse-graining scheme represented neural activity as a continuous field in space and time and was the first neural field theory (see \cite{coombes_waves_2005} for a review).             
Beurle assumed a network of purely excitatory cells with uniform thresholds and refractory period. A neuron becomes active if it receives input from a threshold number of active neurons. If $q$ is the threshold then the rate at which the cells become active at a time $t+\tau$ is
\begin{equation}
F(t+\tau)=R P_{(q-1)} F k
\end{equation}
where $F(x,t)$ is the proportion of cells active in any region per unit time, $k$ is a scale constant, 
$\tau$ is the cell time constant, $R$ is the proportion of cells which are sensitive to input, and
\begin{equation}
P_{(q-1)}=\mathrm{e}^{-\overline{N}} \frac{\overline{N}^{(q-1)}}{(q-1) !}
\end{equation}
is the proportion of cells which have an integrated excitation of $(q -1)$ and are ready to spike with the arrival of one further impulse. 
The mean number of activated neurons given some input $\bar{N}$ obeyed

\begin{equation}
\overline{N}(x, t)=\int_{-\infty}^{0} \int_{-\infty}^{\infty} w(x-x') F(x', t') \alpha(t-t') \mathrm{d} x' \mathrm{d} t'
\end{equation}
where $\alpha(t)$ is the temporal response function and  
$w(x) = b e^{-|x|/a}$ is a spatially dependent connection function inspired by experimental studies \cite{sholl1955organization,uttley1955probability}. 
Again, this was a mean field theory formulation as activity depends only on past activity and not higher moments (cumulants) of the activity.

Beurle analyzed spontaneous random activity and waves. He was interested in waves of activity because of then-recent electroencephalograph studies~\cite{chang1951dendritic,burns1951some,walter1951new,lilly1954surface}.
He showed that his model with minimal structure was capable of supporting plane, spherical and circular waves as well as vortices.  In addition, he was driven by the idea that switching of neural waves may be related to shifting of attention (or other perceptions). 
However, his model also showed that the stationary fixed point is unstable; a slight deviation from the critical point either leads the activity to cease or saturate (due to refractoriness).  Therefore, he erroneously concluded that there is no stable continuous random activity.
Despite the achievements of the theory, the lack of inhibition as well as some technical flaws drove others to look for alternative formulations. Nevertheless, Beurle established most of the concepts that eventually led to the Wilson-Cowan equations. 

The next milestone was Griffith \cite{griffith_field_1963,griffith_field_1965}.
After establishing the role of inhibition in stabilizing brain activity \cite{griffith_stability_1963}, Griffith took a different approach and derived a reaction-diffusion equation for neural activity.
He desired a classical field theory with a field representing the synaptic excitation or input to the cells and a source that quantifies the activity of the cells (e.g.~spiking frequency). He built a phenomenological model by enforcing a set of constraints on the equations.

Griffith adopted a linear inhomogeneous differential equation for $\psi$ of the form:
\begin{equation}
H \psi=\kappa F
\end{equation}
where $H$ is a linear operator and $\kappa$ is a constant.
He defined $F$ as 
\begin{align}
F(x, t) & \equiv f\Big[ \int_{-\infty}^{+\infty} \psi(x, t-\tau) \chi(\tau) d \tau \Big]
\end{align}
where $f$ is a nonlinear activation function and the temporal response function $\chi(\tau)$ obeys causality.
Assuming that $\psi$ is differentiable and quasi-stationary ($\partial^{2} f / \partial \psi^{2} \ll 1 $),  Griffith obtained the following general form for $F$ by Taylor expanding to second order:
\begin{equation}
F \approx f(\psi)+A_{1} \frac{\partial \psi}{\partial t}+A_{2} \frac{\partial^{2} \psi}{\partial t^{2}} \, , \end{equation} 

He then assumed that the field operator $H$ obeyed translational and rotational invariance, which implied
\begin{equation}
H=a+b \frac{\partial}{\partial t}+c \frac{\partial^{2}}{\partial t^{2}} + \nabla^{2} \, .
\label{H}
\end{equation} 
Substituting a field solution ansatz with constant velocity $v$, 
\begin{equation}
\psi(\mathbf{r}, t)=\int F\left(\mathbf{R}, t- p/ v \right) w(|\mathbf{r}-\mathbf{R}|) d \mathbf{R} \, , 
\label{psi}
\end{equation}
where $w$ is a connectivity function,
into equation (\ref{H}), he obtained
\begin{equation}
\begin{aligned} \nabla^{2} \psi=&\frac{1}{4} \beta^{2} \psi+\left(\frac{\beta}{v}-4 \pi A_{1}\right)  \frac{\partial \psi}{\partial t} \\ &+\left(\frac{1}{v^{2}}-4 \pi A_{2}\right) \frac{\partial^{2} \psi}{\partial t^{2}}-4 \pi f(\psi) \end{aligned}
\label{Griffith_general}
\end{equation}
for $w(p)=A p^{-1} e^{-\frac{1}{2}\beta p}$, which Griffith argued was compatible with Sholl's experimental findings \cite{sholl1955organization}. 
Rescaling and assuming a shallow activation function where $\partial f / \partial \psi$ varies slowly, in the limit of $v \to \infty$ equation (\ref{Griffith_general}) can be reduced to 
\begin{equation}
\nabla^{2} \psi= \mu \psi+\dot{\psi} - f(\psi)
\label{RD}
\end{equation}
where $\mu>0$ is a constant and $f$ is also redefined as a general nonlinear function. 
From (\ref{RD}) 
it is apparent that for spatial homogeneity one finds a mean field equation 
\begin{equation}
\dot{\psi} = - \mu \psi + f(\psi)   
\end{equation}
which is the graded-response model (\ref{eqn:Amari_format}) for uniform connectivity, which as shown above is equivalent to the Wilson-Cowan equation.

Griffith showed that Beurle's formalism is equivalent to his if the activation function 
$f(\psi)$ is adjusted appropriately. He originally assumed a linear response function, namely $f(\psi)=f(\psi_c)+b(\psi-\psi_c)$, but argued that a sigmoidal $f$ could have two stable critical points and an unstable fixed point in between that corresponded to Beurle's unstable steady state. Griffith was the first to show the possibility of stable spontaneous activity depending on the gain function. However, he did not account for excitatory and inhibitory synaptic connections explicitly.


\begin{figure}
\begin{center}
\includegraphics[scale=0.28]{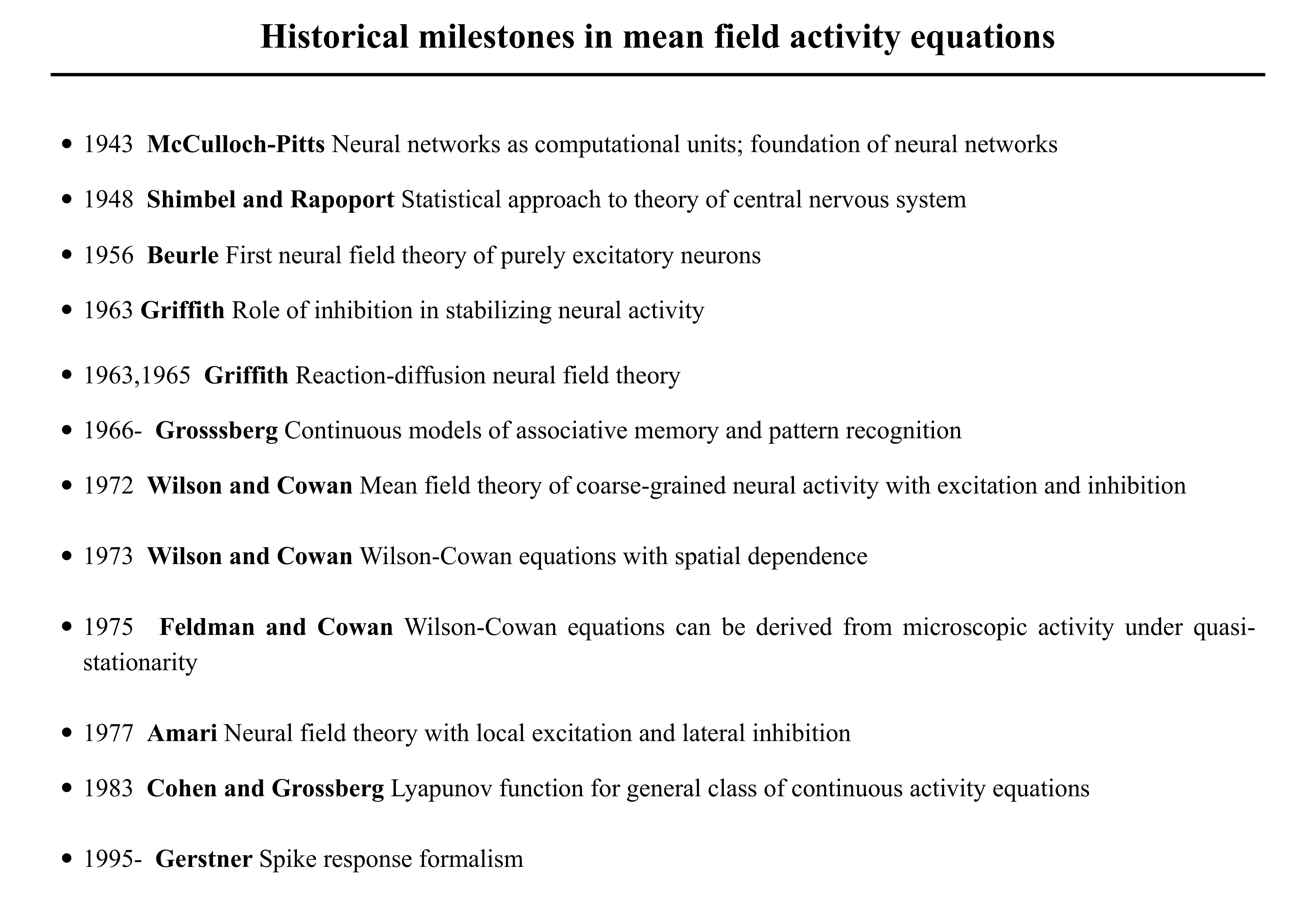}
\caption{Milestones in the development of activity equations.}
\label{Fig3}
\end{center}
\end{figure}


\section{The Wilson and Cowan equations}\label{sec:WC}

By the mid 1960's most of the concepts that would be incorporated into the Wilson-Cowan equation had been proposed although no one had put them together into a cohesive whole.
Wilson and Cowan developed a coarse-grained description of neuronal activity where the distinction between excitatory and inhibitory cells was taken into account explicitly.  
They were motivated by physiological evidence from Hubel and Wiesel \cite{wiesel1963single,hubel1965receptive}, Mountcastle \cite{mountcastle1957modality}, Szentagothai and Lissak \cite{szentagothai1967recent} and 
Colonnier \cite{colonnier1965structural}, which suggested the existence of certain populations of neurons with similar responses to external stimuli. 

In line with Sholl \cite{hosokawa1958sholl} and Beurle \cite{beurle1956properties}, Wilson and Cowan argued that a microscopic (single-neuron) description of neural activity is probably not well suited for understanding higher level functions that entail more complex processes such as sensory processing, memory and learning. 
Using dynamical systems analysis, they showed that their equations exhibit multi-stability and hysteresis, which could serve as a substrate for memory \textbf{\cite{cragg_memory:_1955,fender1967extension}}, and limit cycles, where the frequency of the oscillation is a monotonic function of stimulus intensity. The combination of mathematical tractability and dynamical richness is the reason for the lasting legacy of their equations.

Wilson and Cowan derived their activity equations from first principles using a probabilistic framework on an aggregate of heterogeneous threshold neurons coupled with excitatory and inhibitory synapses with random connectivity by which spatial interactions could be neglected. They extended the model to the case of spatial inhomogeneity the following year \cite{wilson_mathematical_1973}. 

They assumed that that neurons had a sensitive phase where input exceeding a threshold would cause them to fire and a refractory period in which they would not fire regardless of input.  Defining $A_i$ as the proportion of cells of type $i \in \{E,I\}$ (excitatory or inhibitory) active at time $t$, the fraction that are refractory will be $\int_{t-r}^t A_i(t') dt$, 
with refractory period $r$ and the fraction that are sensitive is then $1-\int_{t-r}^t A_i(t') dt$.

The activation function $f(z)$ gives the expected proportion of cells that would respond to a given level of input $z$ if nonrefractory. If this
and the fraction of sensitive neurons are independent, then the updated fraction of active cells would be
\begin{equation}
\left[1-\int_{t-r}^{t} A_i\left(t^{\prime}\right)  dt^{\prime}\right] f_i(z_i) \delta t \, ,
\end{equation}
where 
$$z_i \equiv \int_{\infty}^{t} \alpha\left(t-t^{\prime}\right)\left[W_{iE} A_E\left(t^{\prime}\right)-W_{iI} A_I\left(t^{\prime}\right)+Q_i\left(t^{\prime}\right)\right] dt^{\prime},$$ $Q_i$ is an external input, $\alpha(t-t')$ is a response function governing the time evolution of the spike arrivals, and $W_{jk}$ is the average number of synapses from type $k$ to $j$.
Wilson and Cowan noted that the input to a cell and the sensitive fraction could be correlated, which would violate mean field theory since both involve $A_i$. 
They argued that this correlation could be negligible for highly interconnected networks due to the presence of spatial and temporal fluctuations in the average excitation within the population, and also due to the variability of thresholds supported by experiments  \cite{geisler1958extracranial,verveen1969amplitude,rall1956analysis}.

To estimate the shape of the activation function, they assumed that even if all the neurons receive the same input, any variability in other parameters such as the firing threshold or the number of afferent connections could lead to a probability distribution for the number of spiking neurons. Assuming that the variability mainly stems from the threshold, the expected proportion of neurons that receive at least a threshold excitation (per unit time) would be \footnote{A similar argument holds for a uniform threshold and a distribution of neuronal afferent synapses, for which  $f(x)=\int_{\theta / x(t)}^{\infty} c(w) \mathrm{d}w
$ where $c(w)$ is the probability density for synaptic connections, and the lower bound on the integral is because all neurons with at least $\theta / x(t)$ connections would cross the threshold.}
\begin{equation}
f(x)=\int_{0}^{x(t)} p(\theta) \mathrm{d} \theta \, .
\end{equation}
where $p(\theta)$ is the threshold probability density.
If $p(\theta)$ is unimodal then $f(x)$ would be a sigmoid function.  Later, Feldman and Cowan \cite{feldman_large-scale_1975-1} expanded on that interpretation showing that the activation function in the Wilson-Cowan equations can be regarded as the average of single-neuron activation functions. 

Putting this all together gives
\begin{align}
A_E(t+\tau) = &\bigg [ 1- \int_{t-\tau}^t A_E(t') dt' \bigg]\label{eqn:WC01} \\ & \times f_e\Bigg(\int_{-\infty}^t \alpha(t-t') \Big[W_{EE} A_E(t') - W_{EI} A_I(t') + Q_E(t') \Big] dt' \Bigg)\nonumber\\
A_I(t+\tau) =& \bigg [ 1- \int_{t-\tau}^t A_I(t') dt' \bigg]\label{eqn:WC02} \\
&\times f_I \Bigg(\int_{-\infty}^t \alpha(t-t') \Big[W_{IE} A_E(t') - W_{II} A_I(t') + Q_I(t') \Big] dt' \Bigg) \nonumber
\end{align}

After obtaining the general equations (\ref{eqn:WC01}) and (\ref{eqn:WC02}), Wilson and Cowan derived a set of differential equations that carried the biologically relevant aspects of their general equations. Defining time coarse-grained variables
\begin{equation}
a_i(t) = \frac{1}{s} \int_{t-s}^{t}A_i\left(t^{\prime}\right) \mathrm{d} t^{\prime}
\end{equation}
they argued that if $\alpha(t) \approx 1$ for $0 \le t \le r$ and  decays rapidly for $t > r$,  then 
\begin{align} 
&\int_{t-r}^{t} A\left(t^{\prime}\right) \mathrm{d} t^{\prime} \rightarrow r a(t) \\ &\int_{-\infty}^{t} \alpha\left(t-t^{\prime}\right) A\left(t^{\prime}\right) \mathrm{d} t^{\prime} \rightarrow k a(t) 
\end{align}
where $k$ and $r$ are constants. Inserting into (\ref{eqn:WC01}) and (\ref{eqn:WC02}), Taylor expanding, and rescaling gives

\begin{align}
\tau_E \frac{\mathrm{d} a_E}{\mathrm{d} t}&=-a_E+(1-r a_E) f_{E}\left[ W_{EE} a_E - W_{EI} a_I+ Q_E(t)\right]
\label{eqn:WC1}\\
\tau_I \frac{\mathrm{d} a_I}{\mathrm{d} t}&=-a_I+(1-r a_I) f_{I}\left[ W_{IE} a_E - W_{II}  a_I +  Q_I(t)\right],
\label{eqn:WC2}
\end{align}


After deriving the original mean field equations, Wilson and Cowan considered spatial inhomogeneity. Inspired by Beurle \cite{beurle1956properties}, they extended their original equations to:
\begin{equation}
\begin{aligned} \tau \frac{\partial A_E(\mathbf{x}, t)}{\partial t}=&-A_E(\mathbf{x}, t)+(1-r A_E(\mathbf{x}, t)) \\ & \times f_{E}\left[\int_{\Omega} d \mathbf{x}^{\prime} \rho_{E}(\mathbf{x'}) W_{E E}\left(\mathbf{x}-\mathbf{x}^{\prime}\right) A_E\left(\mathbf{x}^{\prime}, t\right)\right.\\ &-\int_{\Omega} d \mathbf{x}^{\prime} \rho_{I}(\mathbf{x'}) d \mathbf{x}^{\prime} W_{E I}\left(\mathbf{x}-\mathbf{x}^{\prime}\right) A_I\left(\mathbf{x}^{\prime}, t\right)+ I_{E}(\mathbf{x}, t) ]
\label{eqn:WC_space}
\end{aligned} 
\end{equation}
\begin{equation}
\begin{aligned} \tau \frac{\partial A_I(\mathbf{x}, t)}{\partial t}=&-A_I(\mathbf{x}, t)+(1-r A_I(\mathbf{x}, t)) \\ & \times f_{I}\left[\int_{\Omega} d \mathbf{x'} \rho_E(\mathbf{x'})  W_{I E}\left(\mathbf{x}-\mathbf{x}^{\prime}\right) A_E\left(\mathbf{x}^{\prime}, t\right)\right .\\ &-\int_{\Omega} d \mathbf{x'} \rho_{I}(\mathbf{x'}) W_{II}\left(\mathbf{x}-\mathbf{x}^{\prime}\right) A_I\left(\mathbf{x}^{\prime}, t\right)  +I_{I}(\mathbf{x}, t) ] 
\label{eqn:WC_space2}
\end{aligned}
\end{equation}
where the activity variables are 
\begin{equation}
\begin{array}{l}{A_E(\mathbf{x}, t)=\int_{-\infty}^{t} d t^{\prime} \alpha\left(t-t^{\prime}\right) n_{e}\left(\mathbf{x}, t^{\prime}\right)} \\ {A_i(\mathbf{x}, t)=\int_{-\infty}^{t} d t^{\prime} \alpha\left(t-t^{\prime}\right) n_{I}\left(\mathbf{x}, t^{\prime}\right)}\end{array}
\end{equation}
where $\alpha\left(t-t^{\prime}\right)=\alpha_{0} e^{-\left(t-t^{\prime}\right) / \tau}$, $W_{ij} \left(\mathbf{x}-\mathbf{x}^{\prime}\right)$ is the spatially dependent connection weight, $\rho_i(\mathbf{x})$ defines the density of neurons in a small volume around $\mathbf{x}$, and $n_E$ and $n_I$ represent the proportions of excitatory and inhibitory neurons activated per unit time. 
Although Wilson and Cowan developed the model to incorporate spatial dependence of connections, the formulation could go beyond spatial dependency as $\mathbf{x}$ could in general be any abstract quantity. 


\section{Graded Potential models}\label{sec:grossberg}

The artificial intelligence branch of neural networks following McCulloch and Pitts focused initially on the computational capabilities of the binary neuron, e.g. \cite{caianiello1961outline,orbach1962principles,widrow1962generalization}. However, Hartline and Ratliff \cite{hadeler1987stationary} considered the continuous firing rate of single cells in a neural network to successfully model the liminus retina, which could be directly measured. 
This was followed by Grossberg, whose initial goal was to understand how the behavior of an individual can adapt stably in real-time to complex and changing environmental conditions \cite{grossberg1988nonlinear}. 
The core of his approach led to a class of continuous neural networks defined by the nonlinear coupling between activity and synaptic (adaptive) weights. He proved that the computational units of these networks are not individual activities or connections, but are the pattern of these variables (Grossberg,1968b, 1969a,1969b,1970a) \cite{grossberg1968some,grossberg1969learning,grossberg1969some,grossberg1969some,grossberg1988nonlinear}. 

Grossberg first considered model (\ref{eqn:Amari_format}), which he termed the additive STM equation since the equation exhibits bistability as shown by Griffith.  He then added dynamics for the synaptic weights to model long term memory. 
Inspired by the structure of the Hodgkin-Huxley model, Grossberg next derived an equation for neural networks that more closely modeled the shunting dynamics of individual neurons resulting in (\ref{eqn:CG_format}) \cite{grossberg1988nonlinear}. The shunting STM model is approximated by the additive STM model when the activities $s_i$ in (\ref{eqn:CG_format}) are far from saturation.
These networks are capable of performing a rich repertoire of behaviors including content addressable memory and oscillations \cite{cohen_absolute_1983}.
Cohen and Grossberg (1983) proved that under general conditions including symmetric weights, general activity equations possess a Lyapunov function, indicating that all orbits will flow to local minima and can serve as a basis for models of associative memory. 


In 1982, Hopfield using the analogy to the statistical mechanics of spin glasses, showed that a discrete time binary neuron model with symmetric connections has a non-increasing energy function (Lyapunov function) and thus stable attractors that can act as memory states~\cite{hopfield1982neural}.
The foundations of this work had been set in the early 1960's, when several studies demonstrated that (artificial) adaptive networks could perform recognition tasks \cite{rosenblatt1961principles,widrow1960adaptive}. These advances were followed by a number of studies on associative memory and pattern recognition, including those by Grossberg \cite{grossberg1968some,grossberg1969learning,grossberg1969some,grossberg1969some} and Amari \cite{amari_dynamics_1977,amari_field_1983,Potthast2013}.
Little and Shaw \cite{little1974existence,little1978analytic} pointed out the analogy between neural networks and spin glasses.  
Hopfield's contributions \cite{hopfield_computing_1986} attracted a lot of interest from the physics community \cite{amit_spin-glass_1985,sompolinsky1988statistical}, but it is rarely acknowledged that his discovery was a special case of the general Cohen-Grossberg theorem and that there was much work in this area that preceded his.

\section{Spike response formalism}\label{sec:gerstner}

The Wilson-Cowan equations were highly impactful in modeling a number of neural phenomena such as pattern formation, waves, and slow oscillations. However, Feldman and Cowan showed that the Wilson-Cowan equations (\ref{eqn:WC_format}) are only valid for quasi-stationary activity \cite{feldman_large-scale_1975}. Gerstner and van Hemmen \cite{gerstner_universality_1992} also showed that for stationary activity, any single-neuron model can be reduced to a network of graded-response neurons, but this is not true for coherent oscillations \cite{gerstner_universality_1992}.

To address this deficiency, Gerstner sought a general formulation of global dynamics whereby one can systematically estimate the accuracy of rate-based models. To that end, he developed the spike-response formalism, which is a systematic derivation of effective population (macroscopic) dynamics, given that the single neuron (microscopic) dynamics are known \cite{gerstner_time_1995}. His approach echoes the ideas underlying the Wilson-Cowan equations, with a focus on a realistic model of single-neuron dynamics. 

In the spike-response formalism, the membrane potential of a neuron is modeled by the combination of the time dependent refractory response of the neuron to its own activity and the summed responses to the incoming spikes \cite{gerstner2014neuronal,gerstner_universality_1992}. As a result the synaptic potential of a single neuron can be described by the integro-differential equation
\begin{equation}
h_i(t)=  h_i^{\mathrm{ext}}(t)+ \sum_{f=1}^{F} \eta^{\mathrm{refr}}\left(t-t_{i}^{f}\right) 
 + \sum_{j \in \Gamma_{i}} J_{i j} \int_{0}^{\infty} \kappa\left(s, s^{\prime}\right) S_{j}^{(F)}\left(t-s^{\prime}\right) d s^{\prime} 
\end{equation}
where $h_{i}$ is the membrane potential, $\eta^{\mathrm{refr}}$ is the refractory function, $t_i^f$ denotes the spike times for neuron $i$, and $\kappa\left(s, s^{\prime}\right)$ denotes the response kernel in which $s$ is the time that has passed since the last postsynaptic spike represented by $S_j^{(F)}$ \cite{gerstner2014neuronal}. Different models of single-neuron dynamics can be reduced to a spike-response model with appropriate kernel functions \cite{gerstner_universality_1992,gerstner_associative_1992,gerstner_time_1995}. 

In order to find the connection to rate-based models, Gerstner applied mean field theory. He considered a uniform population of neurons for which the activity $A_{p}(t)$ of a single pool $p$ is defined as the proportion of spiking neurons per unit time within the pool. Neurons in a pool are equivalent; connection weights and the response kernel $\kappa(s,s')$ only depend the pool identity. However, the addition of noise, which turned out to be essential in his formulation, does cause variability among spike trains within a single pool. 
To formulate his pool dynamics, Gerstner assumed that the spike trains could be approximated as renewal processes if the synapses are weak $|W_{pq}| \ll 1$. This led to a closed set of mean field equations describing the synaptic input $h(p,t)$, pool activity $A_{p}(t)$ and the firing noise probability imposed by the dynamics. His formalism is suited to model neural network dynamics at arbitrarily short time scales.

In line with previous studies \cite{feldman_large-scale_1975-1,gerstner_universality_1992}, 
Gerstner showed that differential equation activity models, while excellent for modeling asynchronous firing, break down for fast transients and coherent oscillations. He derived a correction to the quasi-stationarity assumption, estimating the error of rate-based models. He showed that in order for the rate-based description to be valid one should have 
$
\langle s \rangle \dot{A}(t)/A(t) \ll 1 \,  
$
where $\langle s \rangle$ denotes the inter-spike interval (inverse of mean firing rate). In addition, he also re-derived the Wilson-Cowan equations by adjusting the appropriate kernels and provided a more rigorous derivation wherein the noise-induced neuron firing probability plays the same role as the sigmoidal activation function in the original Wilson-Cowan derivation.


\section{Mean field theory for known microscopic dynamics}\label{sec:micro}

The Wilson-Cowan equations and variants have successfully modeled and predicted a variety of neural phenomena. However, the question remains as to how quantitatively accurate they are in modeling a neural system with known microscopic neuron and synaptic dynamics, the spike response formalism notwithstanding.  Here, we explicitly derive a mean field theory for the neural activity of a deterministic network of coupled spiking neurons \cite{buice_dynamic_2013,qiu2018finite}. 

Consider a network of $N$ conductance-based neurons:
\begin{align*}
    \tau_V \frac{dV_i}{dt} &= h_V(V_i,m_i,s_i)\\
    \tau_m\frac{dm}{dt} &= h_m(V_i,m_i,s_i)\\
    \tau_s\frac{ds_i}{dt} &= - s_i + \sum_{j=1}^N w_{ij} A(V_j)
\end{align*}
where $V_i$ is the membrane potential of neuron $i$, $m_i$ represents a single or set of auxiliary variables, $s_i$ represents the synaptic input or drive, $h$'s are continuous functions specifying the dynamics of the respective variabls, $\tau$'s are time constants,  $w_{ij}$ is a matrix of synaptic weights between neurons, and $A$ is a function that activates whenever $V_j$ exceeds a threshold indicating an action potential. 
We do not explicitly distinguish between excitatory or inhibitory neurons but this is reflected in the parameter values, which can vary from neuron to neuron, and the synaptic weight matrix (e.g. obey Dale's law).  

If the individual neurons have limit cycle dynamics, as expected for spiking neurons, and the coupling between individual neurons is not excessively strong (i.e. a single spike does not strongly perturb the spiking dynamics, although many can and will), then the neuron dynamics can be reduced to a phase variable around or near the limit cycle \cite{brown_phase_2004}.  The system takes the simpler form of
\begin{align}
    \dot\theta_i &= F_i(\theta_i,s_i)\label{eqn:theta}\\
    \tau_s \dot{s_i} &= - s_i +\frac{1}{N}\sum_j w_{ij} \delta(t-t_s^j)\label{eqn:u}
\end{align}
where $\theta_i$ is the phase of neuron $i$, $F_i$ is the phase velocity, and $t_s^j$ are the spiking times of neuron $j$, which we set to $\theta_j = \pi$.

Our goal is to generate a coarse-grained mean field description of the network specified by (\ref{eqn:theta}) and (\ref{eqn:u}).
We quantify the neuron activity in terms of a density of the neuron phase:
\begin{align}
\eta_i(\theta,t) = \delta(\theta-\theta_i(t))
\end{align}
where $\delta(\cdot)$ is the Dirac delta function.
This allows us to write
\begin{align}
\delta(t-t_s^j) = \dot{\theta_j}\eta(\pi,t) = F_j(\pi,s_j)\eta(\pi,t)
\label{eqn:delta}
\end{align}
which is also the firing rate of the neuron, i.e.~the neuron velocity or flux at $\pi$.  Inserting (\ref{eqn:delta}) into (\ref{eqn:u}) gives
\begin{align}
    \tau_s \dot{s}_i = - s_i +\frac{1}{N}\sum_j w_{ij} F_j(\pi,t)\eta_j(\pi,t)\label{eqn:s}
\end{align}


We obtain the dynamics of $\eta_i$ by
imposing local neuron conservation.  Although $\eta_i$ is not differentiable, we can still formally write
\begin{align}
\partial_t \eta_i(\theta,t) + \partial_\theta \big[ F_i(\theta_i,s) \eta_i(\theta_i,t)\big] = 0
\label{eqn:Klimontovich}
\end{align}
which is called the Klimontovich equation in the kinetic theory of plasmas \cite{ichimaru1973basic,hora1984}. Equations (\ref{eqn:Klimontovich}) and (\ref{eqn:s}) fully specify the spiking neural network dynamics but are no simpler to solve than the original equations.
They need a regularization scheme to make them useful. One approach is to average over a selected population. This was the strategy employed in sections \ref{sec:BWC}, \ref{sec:WC}, and \ref{sec:gerstner} and in previous mean field treatments of networks \cite{strogatz_stability_1991,brunel2000dynamics} where a population average is taken. In the limit of $N\rightarrow\infty$, the phases are assumed to be sufficiently asynchronous so that $\eta$ is differentiable and (\ref{eqn:Klimontovich}) becomes a well behaved partial differential equation. With the addition of Gaussian white noise, it takes the form of a Fokker-Planck equation. 

An alternative is to average over an ensemble of networks, each prepared with a different initial condition drawn from a specified distribution \cite{MBJD2007,2007MichaelPRL,2007MichaelPRE,buice_dynamic_2013,buice2013generalized,MB2010,Michael2013bymf}. 
Taking the ensemble average $\langle\cdot\rangle$ over (\ref{eqn:s}) and (\ref{eqn:Klimontovich}) gives
\begin{align*}
\partial_t \langle \eta_i(\theta,t)& \rangle + \partial_\theta \langle F_i(\theta_i,s_i) \eta_i(\theta_i,t) \rangle = 0\\
\tau_s \langle \dot{s_i} \rangle&= -\langle s_i \rangle +\frac{1}{N}\sum_j w_{ij} F_j(\pi,t)\langle\eta_j(\pi,t)\rangle
\end{align*}
We see that the dynamics of the mean of $\eta_i$ and $s_i$ depend on the higher order moment $\langle F_i(\theta_i,s_i) \eta_i\rangle$. We can construct an equation for this by differentiating $F_i(\theta_i,s_i) \eta_i$, inserting (\ref{eqn:s}) and (\ref{eqn:Klimontovich}) then taking the average again but this will include even higher order moments. Repeating will result in a moment hierarchy that is more complicated than the original equations. 

However, if all higher order moments factorize into products of the means (i.e.~cumulants are zero) then we obtain
the mean field theory of the spiking network (\ref{eqn:theta}) and (\ref{eqn:u}):
\begin{align}
\partial_t &\langle \eta_i(\theta,t) \rangle + \partial_\theta  F_i(\theta_i,\langle s_i\rangle) \langle \eta_i(\theta_i,t) \rangle = 0\label{eqn:mftK}\\
\tau_s \langle \dot{s_i} \rangle&= -\langle s_i \rangle +\frac{1}{N}\sum_j w_{ij} F_j(\pi,t)\langle\eta_j(\pi,t)\rangle\label{eqn:mfts}
\end{align}
which does not match any of the mean field activity equations in Sec \ref{sec:equations}. However, since $F_j(\pi,t)\langle \eta_j(\pi,t) \rangle$ is the ensemble mean firing rate of neuron $j$, then we see that (\ref{eqn:mfts}) has the form of (\ref{eqn:Amari_format}) with time dependent firing rate dynamics given by (\ref{eqn:mftK}). If $\langle \eta_i\rangle$ were to go to steady state then (\ref{eqn:mftK}) with $\partial_t \langle \eta_i(\theta,t)\rangle $ set to zero can be solved to yield
\begin{align*}
    \langle \eta_i(\theta) \rangle = \frac{C_j}{\langle F_j(\theta,s_j)\rangle }
\end{align*}
where $C_j$ is determined by the normalization condition $\int_\theta \eta_j(\theta)d\theta = 1$. We can then obtain  (\ref{eqn:Amari_format}) with gain function
\begin{align*}
    f_j(s_i) = \frac{C_jF_j(\pi,s_j)}{\langle F_j(\theta,s_j)\rangle}
\end{align*}
However, (\ref{eqn:mftK}) is dissipation free and thus will not relax to steady state from all initial conditions without a noise source or finite size effects \cite{2007MichaelPRL,2007MichaelPRE,buice_dynamic_2013}.

The formalism can be applied to any type of neuron and synaptic dynamics. The neuron model (\ref{eqn:theta}) and (\ref{eqn:s}) was chosen for simplicity and also so that the mean field equations would be similar to those of Sec \ref{sec:equations}. The inclusion of higher order time derivatives or nonlinear terms could result in very different mean field equations.
A similar argument to arrive at mean field theory could be applied with the population average, although this will be further complicated by the neuron dependent weight $w_{ij}$, which we will address below. 
No matter what average is used, we still do not know if or when mean field theory is
valid. To answer this question we need to compute the corrections to mean field theory and see when they are small.

\section{Beyond mean field theory}\label{sec:beyond}

Consider again the Klimontovich formulation of the spiking network equations (\ref{eqn:Klimontovich}) and (\ref{eqn:s})
\begin{align}
\mathcal{L}_\eta[\eta_i]&\equiv\partial_t \eta_i(\theta,t) + \partial_\theta F_i(\theta,s_i)\eta_i(\theta,t)-\delta(t-t_0)\eta_i^0(\theta)=0\label{eqn:K2}\\
\mathcal{L}_s[s_i]&\equiv\dot{s}_i(t)+\beta s_i(t)- \frac{\beta}{N}\sum_{j=1}^Nw_{ij}F_j(\pi,t)\eta_j(\pi,t)-\delta(t-t_0)s_i^0=0\label{eqn:s2}
\end{align}
where we have expressed the initial conditions as forcing terms and $\beta = 1/\tau_s$. 

We quantify the ensemble average by 
defining an empirical probability density functional for $\eta_i(\theta,t)$ and $s_i(t)$
as functions constrained to (\ref{eqn:K2}) and (\ref{eqn:s2}), to wit
\begin{align}
{\cal P}[\eta,s | \eta^0,s^0]=
\prod_i \delta\left[\partial_t \eta_i + \partial_\theta F_i(\theta,u_i)\eta_i-\delta(t-t_0)\eta_i^0(\theta)\right]\nonumber\\
\times\delta\left[ \dot{s}_i+\beta s_i-2\beta\frac{1}{N}\sum_jw_{ij}\eta_j(\pi,t)-\delta(t-t_0)s_i^0\right]
\end{align}
${\cal P}[\eta,s | \eta^0,s^0]$ is a Dirac delta functional in function space.
This formal expression can be rendered useful with the functional Fourier transform expression of the Dirac delta functional \cite{Chow2015}:
\begin{align*}
\delta\left[\mathcal{L}_\eta\right]&=\int {\cal D}\tilde{\eta_i} e^{-\int dt d\theta\, \tilde{\eta_i}(\theta,t)\mathcal{L}_\eta[\eta_i(\theta,t)]}\\
\delta\left[\mathcal{L}_s\right]&=\int {\cal D}\tilde{\eta_i} e^{-\int dt \, \tilde{s_i}\mathcal{L}_s[s_i(t)]}
\end{align*}
where $\tilde\eta_i(\theta,t)$ and $\tilde s_i(t)$ are transform variables, called
response fields, and the functional integral is over all possible paths or histories of these functions.
Applying the functional Fourier transform gives the functional or path integral expression
\begin{align}
{\cal P}[\eta,s | \eta^0,s^0]=\int &{\cal D}\tilde{\eta_i}{\cal D}\tilde{s} e^{-\int dt d\theta\, \tilde{\eta_i}\mathcal{L}_\eta[\eta_i] -\sum_i\int dt \tilde{s_i}  \mathcal{L}_s[s_i]}
\end{align}

Assuming fixed $s_i^0$ and a distribution for initial density ${\cal P}[\eta^0]$, we can integrate or marginalize over the initial condition densities to obtain
\begin{align}
{\cal P}[\eta,s]=\int  {\cal D}\eta^0 \,
{\cal P}[\eta,s | \eta^0,s^0]{\cal P}[\eta^0]
\label{eqn:PDF}
\end{align}
If we set $\eta_i^0(\theta)=\delta(\theta-\theta_i(t=0))$, the distribution over initial densities is given by the distribution over the initial phase $\rho_i^0(\theta)$.  Thus we can write $\int{\cal D}\eta^0 P[\eta^0]=\int \prod_i d\theta \rho_i^0(\theta)$.
The initial condition contribution is given by the integral
\begin{align*}
e^{W_0[\tilde{\eta}]}&=\int \prod_i d\theta_i \rho_i^0(\theta_i) e^{\sum_i \tilde{\eta_i}(\theta_i,t_0)}\\
&= \prod_i \int d\theta \rho_i^0(\theta) e^{ \tilde{\eta_i}(\theta,t_0)}\\
&= e^{\sum_i \ln\left( 1- \int d\theta \rho_i^0(\theta) (e^{ \tilde{\eta_i}(\theta,t_0)}-1)\right)}
\end{align*}

Hence, the system given by (\ref{eqn:K2}) and (\ref{eqn:s2}) can be mapped to the distribution functional
\begin{align}
{\cal P}[\eta,s]= \int{\cal D}\eta^0 {\cal D}\tilde{\eta} {\cal D}\tilde{s}e^{-S}
\end{align}
with action $S=S_{\eta}+S_{s}$ given by
\begin{align*}
S_{\eta}&=\sum_i \int_{t_0}^{t_1} dt\int_{-\pi}^\pi d\theta \
\tilde{\eta_i}(\theta,t)[\partial_t\eta_i(\theta,t)+ \partial_{\theta}F_i(\theta,s_i)\eta_i(\theta,t)]\nonumber\\
&+\sum_i\ln\left( 1- \int d\theta\rho_i^0(\theta) (e^{ \tilde{\eta_i}(\theta,t_0)}-1)\right)\\
S_{s}&=\sum_i \int_{t_0}^{t_1} dt \ \tilde{s}_i(t)\left(\dot{s}_i+\beta s_i-\frac{\beta}{N}\sum_j w_{ij}F_j(\pi,s_j)\eta_j(\pi,t)-\delta(t-t_0)s_i^0\right)
\end{align*}
The exponential in the initial data contribution to the action (which is a generating function for a Poisson distribution) can be simplified via the Doi-Peliti-Janssen transformation
\cite{MBJD2007,2007MichaelPRL,2007MichaelPRE,buice_dynamic_2013,buice2013generalized,MB2010,Michael2013bymf}: 
\begin{align*}
\psi_i &= \eta_i\exp(-\tilde{\eta}_i)\\ \tilde{\psi}_i&=\exp(\tilde{\eta_i})-1
\end{align*}
resulting in
\begin{align}
S_{\psi}&=\sum_i\int d\theta dt \tilde{\psi_i}(\theta,t)
[\partial_t\psi_i(\theta,t)+ \partial_{\theta}F_i(\theta,s_i)\psi_i(\theta,t)]\nonumber\\
&+\sum_i\ln\left( 1- \int d\theta \rho_i^0(\theta)  \tilde{\psi_i}(\theta,t_0)\right)\label{eqn:actionK}\\
S_{s}&=\sum_i \int dt \tilde{s}_i(t)\left(\dot{s}_i(t)+\beta s_i-\delta(t-t_0)s_i^0\right.\nonumber\\
&\left.-\frac{\beta}{N}\sum_j w_{ij}F_j(\pi,s_j)(\tilde{\psi}_j(\pi,t)+1)\psi_j(\pi,t)\right)\label{eqn:actions}
\end{align}
where we have not included noncontributing terms that arise after integration by parts.

The probability density functional (\ref{eqn:PDF}) with the action (\ref{eqn:actionK}) and (\ref{eqn:actions}) describes the ensemble statistics for the entire spiking and synaptic drive history of the network. By marginalizing over initial conditions, the functional density is no longer sharply peaked in function space.
The expression cannot be evaluated explicitly but can be computed perturbatively by employing Laplace's method and expanding around the extremum of the integrand. This extremum is given by the point in function space with minimal variation, $\delta S = 0$ (i.e.~the least action principle). The resulting equations of motion, if they exist, are the mean field equations. 

The accuracy of mean field theory can be assessed by computing the variance around the mean field solution using perturbation theory. The inverse network size $1/N$ is a candidate expansion parameter. The action $S$ is comprised of the transformed phase density $\psi_i$, which is not differentiable, but the sum over neurons provides an opportunity to regularize it.
Consider first a homogeneous network, $F_i=F$, and uniform global coupling, $w_{ij} = w$.
Then (\ref{eqn:K2}) and (\ref{eqn:s2}) become
\begin{align}
\partial_t \eta_i(\theta,t) + \partial_\theta F(\theta,s_i)\eta_i(\theta,t)-\delta(t-t_0)\eta_i^0(\theta)=0\label{eqn:K3}\\
\dot{s}_i(t)+\beta s_i(t)- wF(\pi,t)\frac{\beta}{N}\sum_{j=1}^N\eta_j(\pi,t)-\delta(t-t_0)s_i^0=0\label{eqn:s3}
\end{align}
The dynamics of all neurons are identical. We can drop the neuron index and rewrite the action as
\begin{align}
S_{\psi}&=N\int d\theta dt \tilde{\psi}(\theta,t)
[\partial_t\psi(\theta,t)+ \partial_{\theta}F(\theta,s)\psi(\theta,t)]\nonumber\\
&+N\ln\left( 1- \int d\theta \rho^0(\theta)  \tilde{\psi}(\theta,t_0)\right)\label{eqn:actionK3}\\
S_{s}&=N \int dt \tilde{s}(t)\left(\dot{s}(t)+\beta s-\delta(t-t_0)s^0\right.\nonumber\\
&\left.-\frac{\beta}{N}\sum_j wF(\pi,s)(\tilde{\psi}(\pi,t)+1)\psi(\pi,t)\right)\label{eqn:actions3}
\end{align}
Applying the least action principle, which amounts to setting the functional derivatives to zero~\cite{buice2013generalized,qiu2018finite}
returns the mean field theory equations
\begin{align*}
\partial_t \rho(\theta,t) + \partial_\theta F(\theta,s_i) \rho(\theta,t) \rangle = 0\\
\langle \dot{s} \rangle= -\beta\langle s \rangle +\beta w F(\pi,t)\rho(\pi,t)\rangle
\end{align*}
where
\begin{align}
\rho(\theta,t) =  \frac{1}{N}\sum_j \eta_j(\theta,t)
\end{align}
The factor of $N$ in the action indicates that deviations away from mean field will be suppressed as $N\rightarrow\infty$. A perturbative expansion in $1/N$ can be constructed using Laplace's method to compute all higher order cumulants, which go to zero as as $N\rightarrow\infty$~\cite{buice_dynamic_2013,qiu2018finite}.

The same argument can be applied to a heterogeneous network with spatially dependent coupling.  If the coupling function and neuron specific parameters are continuous functions of the neuron index, then mean field theory applies for the density and synaptic drive averaged over a local population of neurons at each location in the network in the limit of $N\rightarrow\infty$ \cite{qiu2018finite}.

\section{Final words}
We have provided a brief and biased view of the ideas behind the Wilson-Cowan equations and their descendants. We show that one can arrive at a similar set of activity equations with first order time derivatives, nonlinear activation (gain) function, and a linear sum over inputs, from various starting points. Despite the fact that the equations are not generally quantitatively accurate, they seem to capture many facets of neural activity and function. They are a set of effective equations that have stood the test of time. Mean field theory is implicitly assumed in all of these equations. Correlations are deliberately ignored to make the equations tractable. However, they are omnipresent in the brain and there is much ongoing effort to account for them. Nonetheless, we predict that forty seven years from now, the Wilson-Cowan equations will still have great utility.

\section{Acknowledgments}
This work was supported by the Intramural Research Program of the NIH, NIDDK.

\bibliographystyle{ieeetr}
\bibliography{MFT.bib}


\end{document}